\documentclass[iop]{emulateapj}

\begin{document}
\shorttitle{BH Magnetospheres}
\shortauthors{Nathanail \&
Contopoulos}

\def\gsim{\mathrel{\raise.5ex\hbox{$>$}\mkern-14mu
             \lower_{\rm 0}.6ex\hbox{$\sim$}}}
+\def\lsim{\mathrel{\raise.3ex\hbox{$<$}\mkern-14mu
             \lower_{\rm 0}.6ex\hbox{$\sim$}}}

\author{
Antonios Nathanail\altaffilmark{1,}\altaffilmark{2,}\altaffilmark{*} and Ioannis Contopoulos\altaffilmark{1}} \affil{\altaffilmark{1} Research Center
for Astronomy and Applied Mathematics, Academy of Athens, Athens
11527, Greece \\ and \\
\altaffilmark{2}Section of Astrophysics, Astronomy and Mechanics, Department of Physics, University of Athens, \\   Panepistimiopolis
Zografos, Athens 15783, Greece}

\title{Black Hole Magnetospheres}

\altaffiltext{*}{antonionitoni@hotmail.com}


\begin{abstract}We investigate the structure of the steady-state
force-free magnetosphere around a Kerr black hole in various
astrophysical settings. The solution $\Psi(r,\theta)$ depends on
the distributions of the magnetic field line angular velocity
$\omega(\Psi)$ and the poloidal electric current $I(\Psi)$. These
are obtained self-consistently as eigenfunctions that allow the
solution to smoothly cross the two singular surfaces of the
problem, the Inner Light Surface (ILS) inside the ergosphere, and
the Outer Light Surface (OLS), which is the generalization of the
pulsar light cylinder. Magnetic field configurations that cross
both singular surfaces (e.g. monopole, paraboloidal) are uniquely
determined. Configurations that cross only one light surface (e.g.
the artificial case of a rotating black hole embedded in a
vertical magnetic field) are degenerate.
We show that, similarly to pulsars, black hole
magnetospheres naturally develop an electric current sheet that
potentially plays a very important role in the dissipation of
black hole rotational energy and in the emission of high-energy
radiation.
\end{abstract}



\section{Introduction}

Several types of powerful high-energy sources exist in the
universe (X-ray binaries, gamma-ray bursts, active galactic
nuclei, etc.). In many of these systems the central engine is
believed to involve a rotating black hole threaded by large scale
astrophysical magnetic fields. It has been shown theoretically
that a rotating black hole can radiate away its available
reducible energy through some kind of generalized Penrose process
(e.g. Lasota et al. 2014). Blandford \& Znajek (1977) (hereafter BZ)
discussed one particular such process where the magnetic field
taps the rotational energy of the black hole and generates
powerful outflows of electromagnetic (Poynting) energy. They
argued that space-time frame-dragging induces an electric field
that is strong enough to `break' the vacuum and establish an
electron-positron force-free magnetosphere. They also obtained the
structure of this magnetosphere for simple monopole and
paraboloidal boundary conditions, and estimated the flux of
electromagnetic energy for low black hole spin parameters.

A few years later, MacDonald \& Thorne~(1982) investigated the
same problem in the `$3+1$' formulation. Their approach gave the
opportunity for astrophysicists not familiar with the geometrical
language of general relativity to enter the field of black hole
magnetospheres and bring with them their expertise from other
areas of astrophysics (e.g. pulsar research). In their formulation
all physical quantities are measured by Zero Angular Momentum
Observers (ZAMOs; also known as local Fiducial Observers or
Fidos). In their Locally Non-Rotating Frame (LNRF) the equations
for the electromagnetic field are very-similar in form to their
respective equations in flat spacetime. Furthermore, Thorne, Price
\& MacDonald~(1986) introduced the so called `membrane paradigm'
where they argued that the key element of the BZ mechanism is the
(stretched) black hole horizon.

Punsly \& Coroniti~(1990) proposed a different perspective based
on a  magnetohydrodynamic (MHD) model.
A few years later Komissarov~(2001, 2004a,b) performed general
relativistic MHD numerical simulations and found that the BZ
monopole solution is aymptotically stable. Nevertheless, he
questioned the horizon theory developed earlier and proposed that
the ergosphere and not the horizon plays the main role in the
electrodynamics. Uzdensky~(2005) exploited the analogy with pulsar
magnetospheres (Contopoulos, Kazanas \& Fendt~1999, hereafter CKF)
and discussed a different possibility, namely the physical
significance of the so called Light Surfaces (LS; see below).
These are surfaces where magnetic field lines as a geometric
construct `rotate' at the speed of light in the same or the
opposite direction with respect to ZAMOs. Quoting his work, `{\em
\ldots one has to consider magnetic field lines that extend from
the event horizon out to infinity. Since these field lines are not
attached to a heavy infinitely conducting disk, their angular
velocity $\omega$ cannot be explicitly prescribed; it becomes just
as undetermined as the poloidal current $I$ they carry.
Fortunately, however, these field lines now have to cross two
light surfaces (the inner one and the outer one). Since each of
these is a singular surface of the force-free Grad-Shafranov
equation, one can impose corresponding regularity conditions on
these two surfaces. Thus, we propose that one should be able to
devise an iterative scheme that uses the two light surface
regularity conditions in a coordinated manner to determine the two
free functions $\omega(\Psi)$ and $I(\Psi)$ simultaneously, as a
part of the overall solution process\ldots We realize of course
that iterating with respect to two functions simultaneously may be
a very difficult task\ldots}.

Contopoulos, Papadopoulos \& Kazanas~(2013, hereafter CPK)
implemented precisely what Uzdensky set out to do a few years
earlier and obtained the monopole solution using the LSs to
self-consistently determine the unknown distributions of the
magnetic field angular velocity and poloidal electric current for
a split monopole force-free configuration. The astrophysical
problem is, however, different since what we actually observe are
jets and not an isotropic outflow. {\em We, therefore, need
something to collimate the jet}. This agent may come in the form
of a paraboloidal boundary along the wind from the surrounding
accretion disk (e.g. Tchekhovskoy, Narayan \& McKinney~2010,
hereafter TNM), or in the form of a background vertical magnetic
field held in place by a disk far from the hole (e.g. Komissarov
\& McKinney~2007; Palenzuela {\em et al.}~2010; Alic {\em et
al.}~2012). In both cases, the collimated morphology of the
outflow is dictated by the specific boundary conditions and not by
the black hole itself.

In this paper we generalize the numerical method of CPK and
improve its stability. Our solutions confirm and generalize the
pioneering results of BZ and of previous time dependent numerical
simulations for spin parameters up to maximal rotation and for
various astrophysical boundary conditions. Our goal is a deeper
understanding of the structure of the rotating black hole
magnetosphere. In \S~2 we re-derive the central equation of our
problem, the general relativistic force-free Grad-Shafranov
equation. In \S~3 we generalize the numerical approach of CPK and
re-derive monopole and paraboloidal solutions. We also investigate
a vertical magnetic field and show that this problem is degenerate
(i.e. there exists an infinity of solutions). An interesting
by-product of our numerical method is that we can also solve the
electro-vacuum problem through the same equation. Finally, in \S~4
we discuss our results and show that, similarly to pulsars, black
hole magnetospheres naturally develop an electric current sheet
that potentially plays a very important role in the dissipation of
black hole rotational energy and in the emission of high-energy
radiation.

\section{Basic Equations}

For completeness of the presentation, we will re-write the
equations of steady-state axisymmetric force-free general
relativistic electrodynamics presented in CPK since, as we argued,
these are the physical conditions that pertain in the black hole
magnetosphere. We assume that the background geometry is Kerr, and
will work in Boyer-Lindquist coordinates $(t, r, \theta, \phi)$.
We will use the  `$3+1$' formulation of MacDonald \&
Thorne~(1982). The metric is:
\begin{equation}
ds^2 = - \alpha^2 dt^2 + \varpi^2(d\phi -\Omega
dt)^2+\frac{\Sigma}{\Delta}dr^2 +\Sigma d\theta^2\ ,
\end{equation}
where,
\[
\alpha = (\Delta \Sigma / A)^{1/2},
\]
\[
\Omega = 2 a   M r /  A\ ,
\]
\[
\varpi =\left( A / \Sigma \right)^{1/2} \sin\theta\ ,
\]
\[
\Sigma = r^2 + a^2 \cos^2\theta\ ,
\]
\[
\Delta = r^2 - 2 M r + a^2\ ,
\]
\[
A = (r^2 + a^2)^2 - a^2 \Delta \sin^2\theta
\]
Here, $M $ is the mass of the black hole
and $a$ its angular momentum ($0\leq a \leq M$),
$\alpha$ is the lapse function, $\Omega$ is the angular
velocity of ZAMOs, and $\varpi$ is the cylindrical radius ($\varpi= r
\sin\theta$ when $a =0$). Through out this
paper we use geometric units where $G=c=1$. $\textbf{e}_i $ is the
spatial coordinate basis and
\begin{equation}
\textbf{e}_{\hat{r}} = (\Delta /\Sigma)^{1/2} \textbf{e}_r ,
~\textbf{e}_{\hat{\theta}} =(1/\Sigma)^{1/2} \textbf{e}_{\theta},
~\textbf{e}_{\hat{\phi}} = \frac{\Sigma}{A \sin\theta}\textbf{e}_{\phi}
\end{equation}
are the unit basis vectors in the ZAMO frame.
We consider steady-state, axisymetric, ideally conducting
magnetospheres where $\textbf{E}\cdot \textbf{B}=0$. Under these
assumptions, Maxwell's equations become
\newcommand{\boldnabla}{\mbox{\boldmath$\nabla$}}
$$\boldnabla \cdot \textbf{B}=0$$
$$\boldnabla \cdot \textbf{E}= 4 \pi \rho_e $$
$$\boldnabla\times (\alpha \textbf{B} ) = 4 \pi \alpha \textbf{J}$$
\begin{equation}
\boldnabla \times (\alpha \textbf{E})= 0\ ,
\end{equation}
and the force-free condition yields
\begin{equation}
\rho_e \textbf{E}+\textbf{J}\times\textbf{B}= 0\ . \label{GRFF}
\end{equation}
Furthermore, the electric ($\textbf{E}$) and the magnetic
($\textbf{B}$) field can be expressed in terms of three scalar
functions $\Psi (r, \theta)$ (the total magnetic flux enclosed in
the circular loop $r=$const., $~\theta =$const. divided by
$2\pi$), $\omega(\Psi)$ (the angular velocity of the magnetic
field lines), and $~I(\Psi)$ (the poloidal electric current
flowing through that loop), as
\begin{equation}
 \textbf{B}(r,\theta)=\frac{1}{\sqrt{A} \sin\theta}\left\lbrace
 \Psi_{,\theta}  \textbf{e}_{\hat{r}}  - \sqrt{\Delta}\Psi_{,r} \textbf{e}_{\hat{\theta}} +\frac{ 2 I \sqrt{\Sigma}}
{\alpha}\textbf{e}_{\hat{\phi}}\right\rbrace \label{B}
\end{equation}
\begin{equation}
 \textbf{E}(r,\theta)=\frac{\Omega -
 \omega}{\alpha \sqrt{\Sigma}}\left\lbrace
 \sqrt{\Delta}\Psi_{,r}  \textbf{e}_{\hat{r}} +
 \Psi_{,\theta} \textbf{e}_{\hat{\theta}} +0
\textbf{e}_{\hat{\phi}}\right\rbrace \label{E}
\end{equation}

Putting all these together, eq.~(\ref{GRFF}) can be transcribed as
$$\left\lbrace \Psi_{,rr} +\frac{1}{\Delta} \Psi_{,\theta \theta} +
\Psi_{,r} \left( \frac{A_{,r}}{A} - \frac{\Sigma_{,r}}{\Sigma}
\right)-
\frac{\Psi_{,\theta}}{\Delta}\frac{\cos\theta}{\sin\theta}\right\rbrace$$

$$ \cdot \left[ 1-\frac{2 M r }{\Sigma}+
\frac{4 M a \omega  r \sin^2\theta}{\Sigma} - \frac{\omega^2 A
\sin^2\theta}{\Sigma}\right]$$

$$+ \left(\frac{2 M r }{\Sigma}-\frac{4 M
a \omega  r \sin^2\theta}{\Sigma}\right)\left(
\frac{A_{,r}}{A}-\frac{1}{r}\right)\Psi_{,r}$$

$$+\left(\frac{\Sigma_{,r}}{\Sigma}-\frac{A_{,r}}{A}\right)\Psi_{,r}$$

$$ -\left( 2 \frac{\cos\theta}{\sin\theta} +
\frac{A_{,\theta}}{A}-\frac{\Sigma_{,\theta}}{\Sigma}\right)
\omega A(\omega  - 2\Omega)
\frac{\Psi_{,\theta}\sin^2\theta}{\Delta\Sigma} $$

$$-2\omega \Omega\varpi^2\frac{\Psi_{,\theta}}{\Delta}
\frac{A_{,\theta}}{A} -2Mr\Sigma_{,\theta}
\frac{\Psi_{,\theta}}{\Delta\Sigma^2}$$

$$-\frac{\omega ' A\sin^2 \theta}{\Sigma}(\omega -
\Omega)\left(
\Psi^2_{,r}+\frac{1}{\Delta}\Psi^2_{,\theta}\right)$$

\begin{equation}
= - \frac{4 \Sigma}{\Delta} I I'\ , \label{GRGS}
\end{equation}
the general relativistic force-free Grad-Shafranov equation.
$(\ldots)_{,i}$ denotes partial differentiation with respect to $i$,
whereas
$(\ldots)'$ denotes differentiation with respect to $\Psi$. Notice
that if we set $M=a=0$ we obtain the standard pulsar equation in
flat spacetime (Sharlemann \& Wagoner~1973). The zeroing of the
term multiplying the second order  derivatives in the previous
equation, namely
\begin{equation}
1-\frac{2 M r }{\Sigma}+\frac{4 M a \omega  r
\sin^2\theta}{\Sigma} - \frac{\omega^2 A \sin^2\theta}{\Sigma} =0
, \label{singularity}
\end{equation}
yields the singular surfaces of the problem, the so called Light
Surfaces (LSs) where
\begin{equation}
\alpha^{-1} (\omega - \Omega) \varpi  =\pm 1
\end{equation}
in units where $c=1$ (MacDonald \& Thorne~1982). One can interpret
the expression in the left hand side of the above equation as the
corotation velocity of magnetic field lines (as a geometric
construct) with respect to ZAMOs. Outside the Outer Light Surface
(OLS), magnetic field lines rotate {\em faster} than the speed of
light with respect to ZAMOs, while inside the Inner Light Surface
(ILS) they also rotate {\em faster} than the speed of light but
{\em in the opposite direction to that of ZAMOs}. The ILS lies
between the horizon and the static limit (the boundary of the
ergosphere), whereas the OLS is asymptotically cylindrical (see
Appendix~B). In the case of pulsars, eq.~(\ref{singularity})
yields the standard pulsar light cylinder $\omega r \sin\theta
=1$. We remind the reader of the `natural radiation condition' at
infinity (energy must flow outwards along all field lines), namely
\begin{equation}
0\leq \omega \leq \Omega_{\rm BH}
\end{equation}
(BZ). $\Omega_{\rm BH}=a/2 M r_{\rm BH}$ is the angular velocity
at the event horizon $r_{\rm BH}=M +\sqrt{M^2 -a^2}$. We would
also like to note here that in an electro-vacuum (no charges, no
currents) one solves the equation
\begin{equation}
\boldnabla\times (\alpha \textbf{B} ) =0 .
\end{equation}
The reader can easily check that the above expression is
equivalent to eq.~(\ref{GRGS}) if we set $\omega=\Omega$ and
$I=0$. In other words, we can solve the electro-vacuum problem by
using the same numerical code that we have developed to solve the
general relativistic Grad-Shafranov equation. It is interesting
that the factor multiplying the second order term of the equation
never becomes zero in electro-vacuum.
\begin{figure*}[t] \centering

\includegraphics[trim=0cm 0cm 0cm 0cm,
clip=true, width=18cm, angle=0]{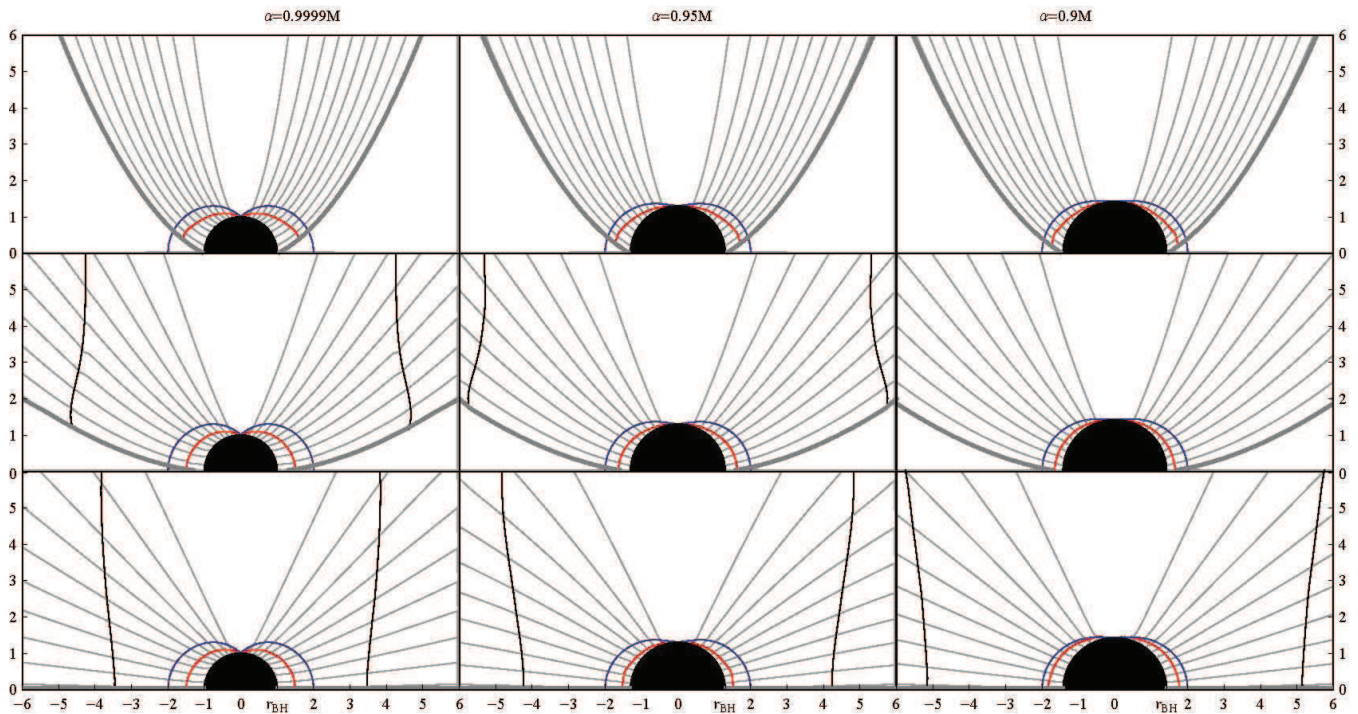}

\caption{Monopole and paraboloidal solutions. $\nu=1$. Top to
bottom: $r_{\rm 0}=1, 10, \infty$. Left to right: $a=0.9999M,
0.95M, 0.9M$. Black semi-circle: the horizon. Red/blue/black
lines: the ILS, the static limit, and the OLS respectively. Where
not seen, the OLS lies outside the region shown. Thick grey line: collimating boundary / equatorial disk. }
\end{figure*}

Equation~(\ref{GRGS}) is a second order elliptic equation for
$\Psi(r,\theta)$ with singular surfaces, and can be solved
numerically with boundary conditions that correspond to various
astrophysical systems. Our numerical method is described in detail
in CPK. We repeat here its main elements which are essentially the
same as the ones implemented by CKF when they obtained the first
solution of the axisymmetric pulsar magnetosphere. Our goal is to
obtain a magnetic field configuration where field lines pass
smoothly through two singular surfaces while at the same time
satisfying the specific boundary conditions of the astrophysical
problem. There are two free functions in the problem,
$\omega(\Psi)$ and $I(\Psi)$, that we can freely adjust in order
to attain our goal. In other words, the solution will break down
at the singular surfaces, {\em except for one particular choice of
$\omega(\Psi)$ and $I(\Psi)$}. Our approach to determine these two
functions is the following: we iteratively evolve the solution
everywhere with a standard relaxation method for elliptic
equations, and every few relaxation steps we adjust the two free
functions by taking into consideration how much the field lines
are `broken' at the two singular surfaces. We find it helpful to
correct $\omega(\Psi)$ at the inner light surface  and $I(\Psi)$
at the outer one. We do not have a theory that will determine the
number of relaxation steps one must wait before updating $\omega$
and $I$, nor how much one must correct them. {\em This is done
empirically with specific correction weights that depend on the
numerical grid resolution}. As in CPK, we changed the radial
variable from $r$ to $R(r)\equiv r/(r+M)$ in order to extend our
numerical integration from the event horizon $r_{\rm BH}$ which
corresponds to $R_{\rm min}\equiv R(r_{\rm BH})$, to some maximum
radial distance $r_{\rm max}$ which corresponds to $R_{\rm
max}=R(r_{\rm max})$. The $\theta$ coordinate extends from the
axis of symmetry ($\theta_{\rm min}=0$) to the equatorial plane
($\theta_{\rm max}=\pi/2$). We implemented a $200\times 64$
numerical grid uniform in $R$ and $\theta$. Notice that this grid
has a very high resolution in $r$ near the event horizon where the
ILS lies, but not as high around the OLS, and in particular at low
$\theta$ and $a$ values. At small $\theta$'s we are, therefore,
obliged to use expansions for $\Psi\propto \sin^2\theta$ (limit of
eq.~\ref{GRGS}), $\omega\rightarrow 0.5 \Omega_{\rm BH}$ (limit of
eq.~\ref{Znajek} below when the field configuration around the
axis is monopolar), and $I\rightarrow 0$.

We have found that the horizon boundary condition is not important
since it is an `inner infinity', and as long as the numerical
relaxation proceeds `smoothly', the horizon regularity condition
\begin{equation}
I(\Psi)=-0.5(\Omega_{\rm BH} - \omega)\frac{\sqrt{A}}{\Sigma}
\Psi_{,\theta} \sin\theta \label{Znajek}
\end{equation}
(Znajek~1977) is automatically satisfied.
The outer (radial infinity) boundary condition is similarly not
important\footnote{In an astrophysical elliptic problem, the
boundary conditions at infinity do not have {\em any effect} on
the solution near the origin, as was the case in CKF. However,
this is not true when one implements boundary conditions in
computational domains of finite spatial extent, as was the case in
Ogura \& Kojima~(2003) where the effect of the boundaries is
easily discerned in their figure~1.} (see Appendix~A).
We discretize all physical quantities and we update
$\Psi(R,\theta)$ through simultaneous overrelaxation with
Chebyshev acceleration (subroutine SOR from Numerical Recipes;
Press, Flannery \& Teukolsky~1986). The final solution does not
depend on our particular choice of initial magnetic field
configuration.

We update the distributions of $\omega(\Psi)$ and $I(\Psi)$ as
follows: At each latitude $\theta$, we check where the singularity
condition (eq.~\ref{singularity}) is satisfied in $r$. At each such
radial position, we extrapolate $\Psi$ inwards from larger $r$
($\Psi(r^+,\theta)$) and outwards from smaller $r$
($\Psi(r^-,\theta)$) using the adjacent three grid points along
$r$. In general, $\Psi(r^+,\theta)$ and $\Psi(r^-,\theta)$ differ.
Then, at the ILS we implement
\begin{eqnarray}
\omega_{\rm new}(\Psi_{\rm new}) & = & \omega_{\rm old}(\Psi_{\rm
new})\nonumber\\
& & -\mu_{\omega}[\Psi(r^+,\theta)- \Psi(r^-,\theta)]\ ,
\end{eqnarray}
whereas at the OLS we implement
\begin{eqnarray}
II'_{\rm new}(\Psi_{\rm new}) & = & II'_{\rm old}(\Psi_{\rm new})
\nonumber\\
& & + \mu_{II'}[\Psi(r^+,\theta)- \Psi(r^-,\theta)]\ ,
\end{eqnarray}
where
\begin{equation}
\Psi_{\rm new} \equiv  0.5[\Psi(r^+,\theta)+ \Psi(r^-,\theta)]
\end{equation}
at each LS. The reasoning here is that we impose weighted
corrections on $\omega$ and $I$ based on the non-smoothness of
$\Psi(r,\theta)$ along the two LS. As we said above, the weight
factors $\mu_{\omega}$ and $\mu_{II'}$ are obtained empirically
and depend on the grid resolution. Notice that we adjust $II'$ and
not directly $I$ since it is $II'$ that appears as a source term
in the right hand side of eq.~(\ref{GRGS}). $I(\Psi)$ is then
obtainable through direct numerical integration, namely
$I(\Psi)=[2\int_0^\Psi II'(\psi){\rm d}\psi]^{1/2}$. This is a
very general procedure that may be applied to any similar singular
equation. The new element with respect to CPK is that once the
values of $\omega(\Psi_{\rm new})$ and $II'(\Psi_{\rm new})$ are
updated, we fit a polynomial of order 5 in $\Psi_{\rm new}$ for
$\omega$, and of order 9 for $II'$. The choice of polynomial order
is determined empirically. The zeroth order term in $\omega$ is
set equal to $0.5\Omega_{\rm BH}$, whereas the zeroth order term
in $II'$ is set equal to zero. All these improvements made the
numerical scheme much more stable than the one implemented in CPK.
\begin{figure*}[t]
\centering
\includegraphics[trim=0cm 0cm 0cm 0cm,
clip=true, width=10cm, angle=0]{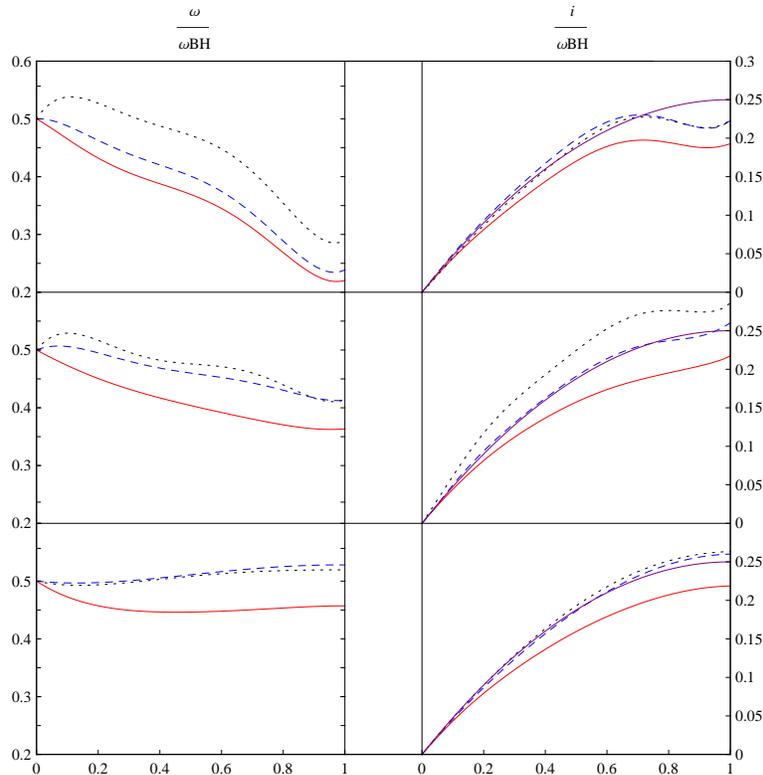} \caption{Left
column: distribution of magnetic field angular velocity
$\omega(\Psi)/\Omega_{\rm BH}$ over $\Psi/\Psi_{\rm BH}$ for
various values of $a$. Right column: distribution of electric
current $I(\Psi)/\Omega_{\rm BH}\Psi_{\rm BH}$.
Solid/dashed/short-dashed lines: $a=0.9999M, 0.95M$ and $0.9M$
respectively. Purple line: Michel's monopole solution in flat
spacetime. Top to bottom: $r_{\rm 0}=1,10, \infty$. $\nu=1$. }
\end{figure*}
Once the relaxation method starts, numerical convergence proceeds
without obstacles. For each value of $a$ and for  particular
boundary conditions, this procedure yields a single set of angular
velocity and electric current distributions and a unique solution
that crosses smoothly the two singular surfaces.

\section{Solutions}

\subsection{Monopole}

As we discussed in CPK, contrary to pulsars, the source of the
black hole magnetic field lies outside the event horizon in the
surrounding distribution of matter (accretion disk, torus, disk
wind, etc.). Therefore, the study of the black hole magnetosphere
necessarily involves the exterior boundary conditions of the
specific astrophysical problem that we are considering, i.e. there
is no such thing as an isolated black hole magnetosphere. In that
respect the pulsar magnetosphere is a `cleaner' problem.

\begin{figure*}[t]
\centering

\includegraphics[trim=0cm 0cm 0cm 0cm,
clip=true, width=18cm, angle=0]{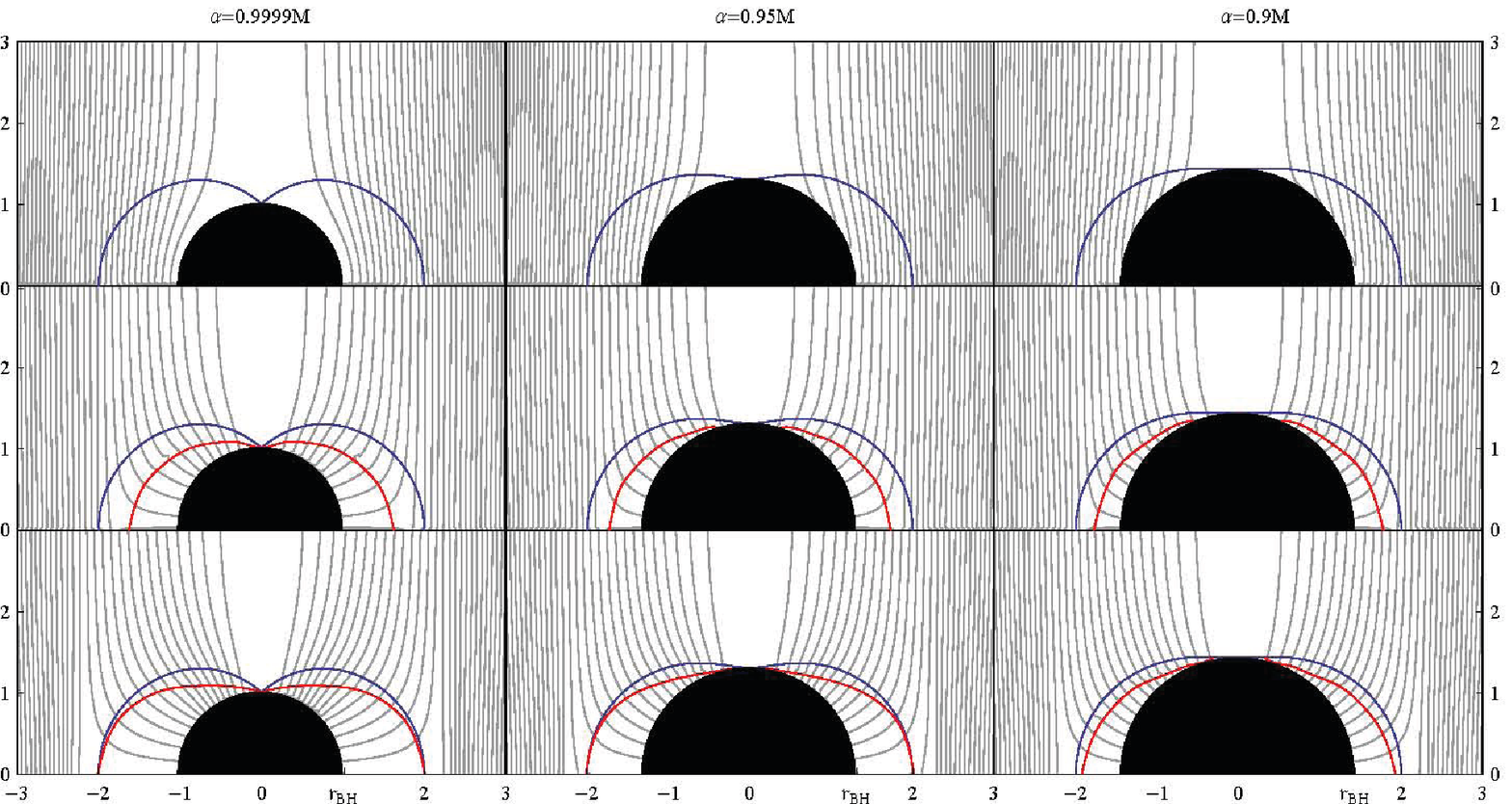}

\caption{Rotating black hole embedded in a vertical magnetic field
for various imposed distributions of $\omega(\Psi)$. Line colors
as in fig.~1. Top row: electrovacuum. Second row: $\omega(\Psi)$
given by eq.~(\ref{omega2}). Notice the presence of field lines
that pass through the ergosphere but do not cross the ILS, thus
they do not carry any electric current, nor any electromagentic
(Poynting) flux. Notice also the development of a poloidal
electric current sheet along the last field line that crosses the
black hole horizon. Third row: $\omega(\Psi)$ given by
eq.~(\ref{omega1}). Notice the suppression of the `Meissner'
effect under force-free conditions.}
\end{figure*}

The simplest magnetized black hole scenario with astrophysical
interest is one where the electric currents supporting its
magnetic field are distributed on a thin equatorial disk that
extends all the way to the horizon. We have found out that the
outer boundary condition is unimportant (as long as the solution
extends smoothly to infinity and fills all space).
We implemented the following boundary conditions on a numerical
grid $(R_{\rm min}\leq R \leq 1, 0\leq\theta\leq \pi/2)$:
\begin{eqnarray}
\Psi(R,\theta=0) & = & 0\ ,\nonumber\\
\Psi(R,\theta=\frac{\pi}{2}) & = & \Psi_{\rm BH}\ ,
\end{eqnarray}
where, $\Psi_{\rm BH}$ is the total magnetic flux threading the
black hole horizon. We initialize our numerical grid with the
Michel~(1982) split monopole, namely
\begin{equation}
\Psi(r,\theta)=\Psi_{\rm BH}(1-\cos\theta)
\end{equation}
with
\begin{equation}
I(\Psi)  =  -0.5 \omega (\Psi) \Psi\left( 2-\frac{\Psi}{\Psi_{\rm
BH}}\right)\ , \ \mbox{and} \label{IMichel}
\end{equation}
\begin{equation}
\omega(\Psi)  =  0.5 \Omega_{\rm BH}\ .
\end{equation}
We also tried different initializations (e.g. paraboloidal) and
the numerical method converged to the same solution. Notice that
eq.~(\ref{IMichel}) is an exact solution for the flat or
Schwarzschild spacetime of a pulsar where one can assume that the
magnetic field angular velocity is constant for all field lines.
For a slowly rotating black hole with small $a$, the solution is
very close to the split monopole as shown by BZ. There is a
practical problem with low $a$'s though: the event horizon is very
close to the boundary of the ergosphere, therefore, the numerical
scheme requires a very high radial resolution in order to treat
the effect of the ILS that lies between the two. At the same time,
the OLS is very far away (remember that it is analogous to the
light cylinder of a slowly rotating pulsar), and therefore, our
numerical procedure is not adequate to treat such a problem. Our
procedure works best for high black hole spin parameters where the
ILS is sufficiently distanced from the event horizon, and the OLS
is not too far away.

As our numerical integration proceeds, both distributions
$\Psi(r,\theta)$ and $\omega(\Psi)$ evolve, and therefore, the
shape and position of the ILS and the OLS also evolve. In other
words, a certain magnetic field line characterized by a certain
value of $\Psi$ will cross each singular surface in a different
place. After about $3000-4000$ iterations the angular velocity and
electric current distributions relax to a steady state that allow
the magnetic flux function $\Psi$ to pass smoothly through the two
LS. The numerical procedure is stable, contrary to CPK where it
was hard to reach an unchanging asymptotic solution. We show
results for three values of $a=0.9999M, 0.95M$ and $0.9M$. As the
black hole become maximally rotating the magnetic flux becomes
concentrated along the axis of symmetry $\theta=0$. Our solutions
directly reproduce and confirm the results of previous time
dependent numerical simulations (e.g. TNM).

The obtained electric current distribution is implemented
physically through an outflow of electrons beyond the OLS that is
connected to an inflow of positrons inside the ILS all the way to
the event horizon. We will not discuss the origin of these
particles and we can only refer the interested reader to
discussions of the wind generation zone in the literature
(Komissarov 2004b, Globus \& Levinson 2013). What is most
interesting, though, is that, similarly to pulsars, in this model
too $I(\Psi_{\rm BH})\neq 0$, and as a result, the poloidal
electric circuit needs to close in the form of a current sheet
along the last open magnetic field line that crosses the black
hole horizon. Therefore, the magnetospheric current sheet seems to
be a generic feature of black hole magnetospheres, and as is the
case with pulsars, this is expected to be associated with the
origin of high energy radiation (see \S~4 ).

Another obvious result is that equatorial boundary conditions do
not produce any collimated energy outflow that would resemble an
astrophysical jet. In other words, as in pulsars, {\em black hole
magnetospheres do not naturally produce jets}. Nevertheless, jets
are observed in nature, and therefore, what is needed is some
collimating agent that would restrict the lateral expansion of the
magnetosphere. This can be found in the form of a disk wind or a
thick torus configuration surrounding the magnetic flux that
threads the black hole horizon.

\subsection{Paraboloidal}

Following TNM we considered a paraboloidal wall described by
\begin{equation}
1-\cos\theta_{\rm wall} = \left(\frac{ r + r_{\rm 0} }{r_{\rm BH}
+r_{\rm 0}}\right)^{-\nu}. \label{wall}
\end{equation}
as a boundary condition that would yield collimated solutions.
$\nu$ is just a shape parameter. Notice that the smaller $r_{\rm
0}$ the more collimated the boundary. As in the monopolar case,
the wall intersects the event horizon at the equator. We
implemented the following boundary conditions on a numerical grid
$(R_{\rm min}\leq R \leq R_{\rm max}, 0\leq\theta\leq \pi/2)$:
\begin{eqnarray}
\Psi(R,\theta=0) & = & 0\ ,\nonumber\\
\Psi(R,\theta\geq \theta_{\rm wall}(r)) & = & \Psi_{\rm BH}\ ,
\nonumber\\
\Psi_{,R}(R=R_{\rm max},\theta<\theta_{\rm wall}(r_{\rm max}) & =
& 0\ .
\end{eqnarray}
The reason we cannot really extend our numerical integration to
$R=1$ is that we used the same numerical grid that we used in the
monopole case, and solved for $\Psi$ only for angles $\theta\leq
\theta_{\rm wall}(r)$. The grid resolution becomes worse and worse
as we move higher up along the paraboloidal wall, and at some
distance that corresponds to about $r_{\rm max}=13M$, our
numerical iteration at the OLS breaks down. A better numerical
approach would have been to re-write our equations in a
paraboloidal grid $(0\leq R\leq 1,0\leq \Theta\leq 1)$ where
$\Theta\equiv \theta/\theta_{\rm wall}(r)$. We initialize our
numerical grid with:
  \begin{equation}
\Psi(r,\theta) = \Psi_{\rm BH}\left( \frac{r +r_{\rm 0}}{r_{\rm
BH} + r_{\rm 0}}\right)^{\nu}(1- \cos \theta).
\end{equation}
The split monopole is a special case of the above configuration
with $r_{\rm 0}=\infty$ or $\nu=0$.

Our numerical integration showed that the angular velocity changes
dramatically from the monopole one, going from $0.5 \Omega_{\rm
BH}$ on the axis to about $0.3\Omega_{\rm BH}$ near the wall. One
can see in the figures how the outer light surface deforms because
of the change in the distribution of the magnetic field angular
velocity. As described previously, all paraboloidal solutions
contain a return poloidal electric current sheet that now flows
along the wall that collimates the jet. Similarly to the monopole
configuration, this too is expected to be associated with the
origin of high energy radiation from a collimated black hole
magnetosphere.

\subsection{Vertical magnetic field}

We next consider the case of a black hole embedded in a vertical
magnetic field. This is obviously an artificial scenario that has
nonetheless been considered by many previous research groups
(Palenzuela {\em et al.} 2010, Komissarov \& McKinney~2007 ). The
electro-vacuum case has been first studied by Wald~(1974), where
it was shown that a maximally rotating black hole expels the
magnetic flux from the vicinity of the event horizon (King, Lasota
\& Kundt~1975). This has been named the black hole 'Meissner
effect' (in analogy to superconductors) in the literature
(Komissarov \& McKinney~2007). We too can easily reproduce the
electro-vacuum solutions. For $a=0.9999M$ the magnetic flux is
almost totally expelled from the event horizon, but as we move to
lower spin parameter values more and more flux passes through the
event horizon.

The force-free case is more interesting. We implemented the
following boundary conditions on a numerical grid $(R_{\rm
min}\leq R \leq 1, 0\leq\theta\leq \pi/2)$:
\[
\Psi(R,\theta=0) = 0\ ,
\]
\[
\Psi(R,\theta) = \Psi_{\rm max}\left(\frac{r  \sin \theta}{r_{\rm
BH}}\right)^2\ \mbox{for}\ r\sin\theta\geq 4M\ ,
\]
\begin{equation}
\Psi_{,R}(R=1,\theta) = 0\ ,
\end{equation}
where $\Psi_{\rm max}$ is some canonical value for the magnetic
flux. Here too, we used the same numerical grid that we used in
the monopole case, and solved for $\Psi$ only for cylindrical
radii $r\sin\theta\leq 4M$. We initialize our numerical grid with
a uniform vertical field
\begin{equation}
\Psi(r,\theta)= \Psi_{\rm max}\left(\frac{r  \sin
\theta}{r_{\rm BH}}\right)^2\ ,
\end{equation}
One can easily see that there is a problem. Some magnetic
field lines cross the ILS, but none crosses both the ILS and OLS.
Therefore, we cannot implement the previous numerical procedure
where both $\omega(\Psi)$ and $I(\Psi)$ are determined through the
condition of smooth crossing of both LSs. We thus implement a
different approach:

It is well known that no field lines can cross the same LS twice
(Gralla \& Jacobson~2014). This is exactly analogous to what
happens in the pulsar magnetosphere where closed stellar field
lines that cross the light cylinder open up to infinity because
matter tied to those field lines cannot corrotate. The same
applies here. Field lines that cross the ILS open from inside and
are stretched all the way to `inner infinity', i.e. the event
horizon. We do not know a priori which field lines will cross the
ILS, and which will not. As we evolve our numerical relaxation,
the system pulls some flux toward the event horizon, and an ILS
forms. The lines that do not cross the ILS cross the
equator vertically ($\Psi_{,\theta} (r,\pi/2)=0$). Obviously,
because of north-south symmetry, no electric current can flow
along such field lines, and therefore $II'(\Psi)$ must equal zero
along them. The ones that do cross the ILS, though, are brought to
the event horizon through an equatorial boundary condition of
horizontal field along the equator between the horizon and the
edge of the ILS inside the ergosphere. This obviously corresponds
to an electric current sheet which, in order to guarantee electric
circuit closure, continues along the last field line that crosses
the horizon to infinity. This must be implemented manually as in
CKF, otherwise the force-free Grad-Shafranov equation treats it
just as a discontinuity in $II'$ and entirely misses its
effect\footnote{In practice, we set $I(\Psi)=I(\Psi_{\rm BH})\exp
[-(\Psi-\Psi_{\rm BH})^2/2\sigma^2]$ for $\Psi>\Psi_{\rm BH}$, and
$\sigma$ is an arbitrary parameter that characterizes the
effective current sheet width. $\sigma$ must be chosen as small as
practically possible given a particular numerical grid resolution.
Notice that if we do not implement this current sheet, the
solution will be different, as was the case in Ogura \&
Kojima~2003.}. The presence of this current sheet is evident
through the discontinuity in the azimuthal component of the
magnetic field $B_\phi$, which imposes an equivalent discontinuity
in the poloidal field (in order to guarantee continuity of
$B^2-E^2$) across the last field line that crosses the black hole
horizon. The poloidal field discontinuity can be discerned in the
solutions presented in Figure~3.
As we said the solution passes only from one LS, so we
cannot determine both $\omega(\Psi)$ and $I(\Psi)$. In other
words, {\em the solution is degenerate}.

In practice, what we did was to {\em arbitrarily specify} a
certain distribution for the angular velocity $\omega(\Psi)$ and
then find the unique electric current distribution $I(\Psi)$ that
allows a smooth solution through the ILS. We obtained typical
results for
\begin{eqnarray}
\omega & = & 0.5 \Omega_{\rm BH}
\cos^2\left(\frac{\pi}{2}\frac{\Psi}{\Psi_{\rm BH}}
\right)\ ,\\
\omega & = &  0.5 \Omega_{\rm BH} \left(1 - \frac{\Psi}{\Psi_{\rm
BH}}\right)^2 \label{omega1}
\end{eqnarray}
etc., and $\omega=0$ for $\Psi>\Psi_{\rm BH}$. $\Psi_{\rm BH}$,
the maximum magnetic flux that threads the horizon, evolves as the
numerical relaxation progresses. In these cases the ILS meets the
boundary of the ergosphere at the equator, and all field lines
entering the ergosphere are rotating and are filled with electric
current that contributes to the total electromagnetic output of
the solution. This choice is arbitrary. We realized that we don't
need to require $\omega(0)=0.5\Omega_{\rm BH}$ along the axis when
the field configuration {\em is not} monopolar there. We tried
arbitrary values of $\omega(0$) (e.g. $0.8\Omega_{\rm
BH}$, $0.3\Omega_{\rm BH}$) and still obtained smooth solutions,
each one for a different current distribution. The same applies on
the equator. The angular velocity does not  have to go to zero
there. We thus also solved the equation for different
distributions like
\begin{equation}
\omega=\Omega_{\rm BH}\left(0.5 - \frac{\Psi}{\Psi_{\rm BH}}
0.25\right) \label{omega2}
\end{equation}
etc. For the latter distribution, $\omega$ reaches
$0.25\Omega_{\rm BH}$ at the equator (other values work equally
well), and therefore, the ILS ends inside ergosphere. Field lines
that do not cross the ILS and pass through the ergosphere do
rotate but they don't carry any electric current ($I(\Psi)=0$) so
they don't take part in the electromagnetic output of the
solution. The angular velocity of these field lines is set to fall
gradually to zero with radial distance, in a way that none of them
accidentally crosses any LS. Moreover, these field lines cross the
equator vertically, and one can easily check (using eqs.~\ref{B},
\ref{E} and \ref{singularity}) that, along the equator,
$B^2-E^2=(B^\theta)^2-(E^r)^2>0$ everywhere outside the ILS, and
$B^2-E^2=0$ at the point where the ILS crosses the equator. This
supports our previous conclusion that no field lines that cross
the ILS can cross the equator vertically inside the ILS ($B^2-E^2$
would become negative there). Instead, such field lines are
stretched all the way to the event horizon, forming the equatorial
current sheet that we described above. Therefore, {\em energy and
angular momentum are extracted along all magnetic field lines that
penetrate the ILS, even when the ILS lies completely inside the
ergosphere}, as is the case shown in the middle row of Figure~3.
This result generalizes the discussion in Section 6.1.2 of
Komissarov~(2004a).

There is still interest in the so called `Meissner effect' because
of the possibility that it `{\em  could quench jet power at high
spins} (Pena~2014). We believe that in a force-free plasma filled
environment (i.e. electron-positron plasma) no Meissner effect
will ever occur. As we discussed previously, if somehow magnetic
flux is brought close to the horizon  field lines passing the ILS
twice will open from the inside and will be naturally stretched
all the way to the horizon. Therefore, for a high spin black hole,
magnetic flux {\em will not be expelled from the horizon}.
Instead, the system {\em will bring flux to the horizon}. This
process is fully supported by our solutions of a maximally
rotating black hole embedded in a force-free vertical magnetic
field.

\section{Discussion}

Blandford \& Znajek (1977) estimated that typical astrophyscial
magnetic fields will naturally break down the vacuum around
astrophysical black holes and will establish a force-free
magnetosphere. In the last decade general relativistic
magnetohydrodynamic simulations were performed to explore the BZ
mechanism and its relation with jet observations (McKinney \&
Gammie 2004, TNM, Sadowski {\em et. al.}~2013). What we learn from
these simulations is that in the region close to the polar axis
and near the black hole the plasma density becomes so low that
artificial corrections must be implemented (matter density and
energy `floors') in order to keep the simulations running. This
suggests that the system wants to impose force-free conditions. In
other words, force-free is a good approximation in these regions.

For the case of a black hole embedded in a vertical magnetic field
we have shown  that we can have infinitely many solutions for each
value of $a$. A particular limit of that problem, one with a
uniform field at infinity, has been addressed by several research
groups through time dependent simulations (Komissarov \& McKinney
2007; Palenzuela {\em et al.}~2010; Alic {\em et al.}~2012). These
simulations are shown to converge to a unique (for each code)
distribution for the angular velocity and electric current. In
this particular case, the condition of a uniform vertical field at
infinity imposes one extra condition between $\omega$ and $I$,
namely $I=-\omega\Psi$ (eq.~(\ref{Iomega}); Appendix~A), and
therefore, $\omega$ and $I$ are uniquely defined. Furthermore,
$\omega(\Psi)=0$ along field lines that do not thread the black
hole horizon, since $I(\Psi)=0$ along those field lines. As a
result, both $\omega(\Psi)$ and $I(\Psi)$ must reach zero along
the last field line $\Psi_{\rm BH}$ that threads the horizon, and
therefore, the ILS reaches the static limit along the equator, and
the poloidal electric current sheet disappears. We plan to address
this problem in a future work, were we will also perform a
systematic comparison between our results and the results of
previous numerical simulations.

Another reasonable  question is what is the astrophysical
significance (if any) of such solutions (a black hole embedded in
a vertical magnetic field). Magnetic field lines are generated by
electric currents in a surrounding distribution of matter, and
therefore, the vertical extent of the vertical field region cannot
extend beyond `a few times' the radial extent of the inner edge of
that distribution of matter. Magnetic field lines must close, and
when closure is taken into consideration, an OLS will develop
which will uniquely constrain the solution. Another problem arises
when people consider vertical magnetic fields in the problem of
two merging black holes. In that case, the spiralling black holes
empty a region of matter one or two orders of magnitude more
extended than their respective gravitational radii.
Correspondingly, the magnetic field that can be held there by the
surrounding matter distribution is at least one order of magnitude
smaller than the canonical value that would correspond to a field
held by an accretion disk that extends down to the ISCO of the
black hole. In that case, the electromagnetic power is in reality
at least two orders of magnitude smaller than estimates obtained
in the literature (see Lyutikov~2010 for details). We conclude
that vertical magnetic field configurations are artificial and
such solutions do not correspond to real astrophysical jets.

A very important element in all our solutions is the poloidal
current sheet that is naturally formed,  where the electric
circuit closes. The current sheet lies along the equatorial thin
disk in the monopole solutions, and along the boundary wall as we
pass to the paraboloidal solutions. As in pulsars, high energy
radiation is expected from reconnection processes that result in
particle acceleration along these current sheets (Lyubarsky \& Kirk 2001;
Li, Spitkovsky \& Tchekhovskoy 2012; Kalapotharakos {\em et al.}~2012).
 For the monopole and paraboloidal solutions this
implies that high energy radiation may not be coming along the
axis of rotation but in a direction orthogonal to it, as in the
orthogonal gamma-ray burst model of Contopoulos, Nathanail \&
Pugliese~(2014).

We are also interested in the power output of our solutions which
we can directly calculate once we obtain the corresponding
distribution of $\omega(\Psi)$ and $I(\Psi)$ as
\begin{equation}
P=2 \int^{\Psi_{\rm BH}}_0 \omega (\psi) I(\psi)d\psi\ .
\end{equation}
In all cases, $P\approx \Omega_{\rm BH}^2 \Psi_{\rm BH}^2$. One
sees directly that the total magnetic flux $\Psi_{\rm BH}$
accumulated through the horizon and the black hole angular
velocity $\Omega_{\rm BH}$ contribute together to the power output
of the black hole. Most people focus only on the role of the black
hole spin, but the issue of the flux accumulation remains equally
(and probably even more) important. This straightforward result
has sometimes been overlooked in the debate on whether a jet
power-black hole spin association is observable in X-ray binaries
black hole jets (Steiner, Narayan \& McClintock~2013, Russell {\em
et al.}~2013). It may also account for the famous AGN radio loud /
radio quiet dichotomy (Sikora {\em et al.}~2007). The efficiency
of the flux accumulation close to the black hole may be related
either to the efficiency of flux advection (e.g. Sikora \&
Begelman~2013), or our favorite, the in situ flux generation by
the Poynting-Robertson Cosmic Battery effect (Contopoulos \&
Kazanas~1998).

Finally, a comment on the extent of the accretion disk is in
order. In all our monopole and paraboloidal solutions the
accretion disk extends all the way to the horizon. This may not be
true in reality. The disk holding the magnetic flux may end
several gravitational radii away from the black hole. As an
extension of our work we would like to investigate such solutions.
We expect that, because the accretion disk will stop before
reaching the horizon, magnetic field lines will cross the ILS
twice, thus they will open from inside and will be stretched all
the way to the `inner infinity', the event horizon. Another
interesting scenario that we would like to investigate is one
where the accretion disk is starting to disperse leaving behind a
rotating magnetospheric charge of finite radial extent. This will
correspond to a ring current that will generate its own dipole
magnetic field (as in Lyutikov~2012). At large distances, such
configurations will be similar to pulsar magnetospheres.


\acknowledgements We thank the referee, Pr. Serguei Komissarov,
for his constructive comments and criticism. This work was supported by
the General Secretariat for Research and Technology of Greece and
the European Social Fund in the framework of Action `Excellence'.

{}

\section*{Appendix~A: The regularity conditions at `infinity'}

The horizon {\em regularity} condition was first derived by Znajek
(1977) and was used as a {\em boundary} condition in the
perturbative solution obtained in Blandford \& Znajek (1977).
MacDonald \& Thorne~(1982) derived it in their formalism and
applied it in the `membrane paradigm' (Thorne {\em et al.}~1986).

Our understanding of the event horizon is that it is an `inner
infinity' similar to the normal (`outer') infinity. Any
electromagnetic field configuration that is generated by a finite
(spatially) distribution of electric currents and extends from the
horizon to infinity $r\rightarrow\infty$ with both $r\sin\theta$
and $r\cos\theta\rightarrow\infty$ along every field line
satisfies the radiation condition $E=B$ there (this is not the
case for a vertical magnetic field), with $\textbf{E}\times
\textbf{B}$ pointing in the direction of `infinity'. Under
steady-state axisymmetric force-free conditions in particular,
\begin{equation}
E^\theta = \pm B^\phi\ \mbox{at `infinity'} \label{infinity}
\end{equation}
(from eqs.~\ref{B} and \ref{E}). At the outer infinity the plus
sign applies and the above equation yields
\begin{equation}
I(\Psi)=-0.5\omega(\Psi) \Psi_{,\theta} \sin\theta\ .
\label{newZnajek}
\end{equation}
It is straightforward to show that, when we multiply
eq.~(\ref{newZnajek}) with its $\Psi$ derivative, we obtain
\begin{equation}
4II'=\left(\omega^2 \Psi_{,\theta \theta} +\omega^2
\frac{\cos\theta}{\sin\theta}\Psi_{,\theta} +\omega \omega'
\Psi^2_{,\theta}\right)\sin^2\theta\ , \label{newZnajek2}
\end{equation}
which is just eq.~(\ref{GRGS}) in the limit $r\rightarrow\infty$
when $r^2\Psi_{,rr}$ and $r\Psi_{,r}$ are much smaller than both
$\Psi_{,\theta\theta}$ and $\Psi_{,\theta}$ (along any field line
that extends to infinity, $\Psi\rightarrow \mbox{const.}$ and
$r\Psi_{,r}\rightarrow 0$, because otherwise, if
$r\Psi_{,r}\rightarrow a\neq 0$, that would imply $\Psi\rightarrow
a\ln r+\mbox{const.}$, and therefore $a=0$). Similarly, at the
inner infinity (the black hole horizon) the minus sign applies and
eq.~(\ref{infinity}) yields
\begin{equation}
I(\Psi)=-0.5(\Omega_{\rm BH} -
\omega(\Psi))\frac{\sqrt{A}}{\Sigma} \Psi_{,\theta} \sin\theta\ ,
\label{Znajek2}
\end{equation}
the Znajek condition (eq.~\ref{Znajek}). Here too it is
straightforward (albeit more tedious) to show that, when we
multiply eq.~(\ref{Znajek2}) with its $\Psi$ derivative, we obtain
\begin{equation}
4 II'=\left( (\omega-\Omega_{\rm BH})^2\Psi_{,\theta \theta}
+(\omega-\Omega_{\rm BH})^2\frac{\cos\theta}{\sin\theta}
\Psi_{,\theta}
-\frac{\Sigma_{,\theta}}{\Sigma}\Psi_{,\theta}+\omega'
(\omega-\Omega_{\rm
BH})\Psi^2_{,\theta}\right)\frac{A}{\Sigma^2}\sin^2\theta\ ,
\label{Znajek22}
\end{equation}
which too is just eq.~(\ref{GRGS}) in the limit $r\rightarrow
r_{\rm BH}$ with $\Delta\rightarrow 0$.

We have thus shown that the two `infinity' conditions do not teach
us anything about the $\omega(\Psi)$ nor the $I(\Psi)$ of
astrophysical solutions. These are determined from the two LSs.
Any `smooth' astrophysical solution that extends all the way from
the horizon to infinity with both $r\sin\theta$ and
$r\cos\theta\rightarrow\infty$ along every field line adjusts
$\Psi_{,\theta}$ so as to satisfy both equations. We did check
that this is indeed the case in all our solutions to within 0.5\%.
Obviously, the other boundary conditions (along the equator, or
the surrounding collimating boundary) and the requirement that the
magnetosphere fills all space are also important in determining
the solution that corresponds to the particular astrophysical
problem under consideration. The same holds true in pulsars where
the solution there is fully determined by the light
cylinder\footnote{Pulsars have one extra boundary condition,
namely that their magnetosphere originates on the stellar
surface.}.

Notice that the vertical field configuration also satisfies
eqs.~(\ref{Znajek2}) \& (\ref{Znajek22}) {\em but not}
eqs.~(\ref{newZnajek}) \& (\ref{newZnajek2}). At infinity, the
problem becomes one of special relativity, and the vertical field
configuration is better described through the generalized pulsar
equation in cylindrical coordinates $\varpi\equiv r\sin\theta$ and
$z\equiv r\cos\theta$, namely
\begin{equation}
4II'=\omega^2\varpi^2  (\Psi_{,\varpi
\varpi}+\Psi_{,\varpi}/\varpi ) +\omega \omega'\varpi^2
\Psi_{,\varpi}^2-(\Psi_{,\varpi \varpi}-\Psi_{,\varpi}/\varpi)
\label{PE}
\end{equation}
(eq.~10, Contopoulos~2005). It is interesting that, in the
particular limit of a uniform vertical magnetic field at infinity,
$\Psi\propto \varpi^2$ and eq.~(\ref{PE}) yields $II'=\omega^2
\Psi+\omega\omega'\Psi^2$ and therefore
\begin{equation}
I=-\omega\Psi\ . \label{Iomega}
\end{equation}

\section*{Appendix~B: Light surfaces}

We generated Figure~4 by solving eq.~(\ref{singularity}) for
constant $\omega$ because we want the reader to appreciate how the
Light Surfaces (LSs) change shape as the spin of the black hole
and the magnetic field angular velocity change. We are not the
first to discuss the role of LSs. They were introduced from the
very beginning by Blandford \& Znajek (1977), and a nice
mathematical description can be found in Komissarov (2004a). For
$\omega=0$ the ILS meets the static limit, the boundary of the
ergosphere. For $\omega= \Omega_{\rm BH}$ the inner light surface
meets the event horizon.

\begin{figure*}[b]
\centering

\includegraphics[trim=0cm 0cm 0cm 0cm,
clip=true, width=15cm, angle=0]{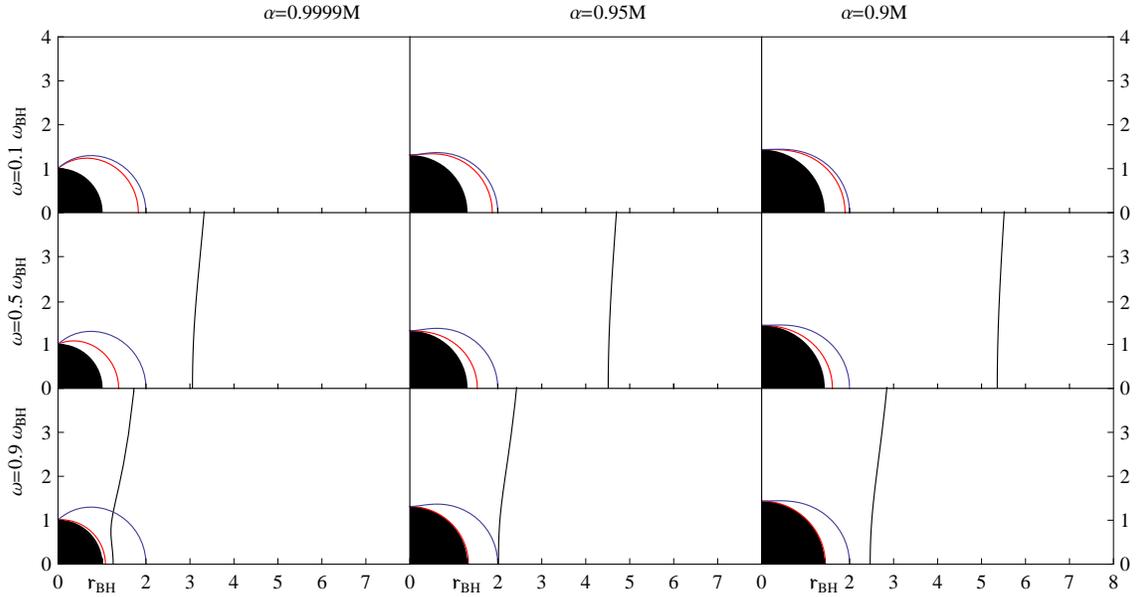}

\caption{Light Surfaces for various black hole spin parameters $a$
and magnetic field angular velocities $\omega$. Line colors as in
fig.~1.}
\end{figure*}


\begin{thebibliography}{}

\bibitem{AMRZJ} Alic, D., Moesta, P., Rezzolla, L., Zanotti, O. \&
Jaramillo, J. L. 2012, ApJ, {\bf 754}, 36
\bibitem{BZ77} Blandford, R. D. \& Znajek, R. L. 1977, MNRAS, {\bf 179},
433 (BZ77)
\bibitem{C05} Contopoulos, I. 2005, A\& A, {\bf 442}, 579
\bibitem{CK98} Contopoulos, I. \& Kazanas, D. 1998, ApJ, {\bf 508}, 859
\bibitem{CKF99} Contopoulos, I., Kazanas, D. \& Fendt, C. 1999, ApJ, {\bf 511}, 351
\bibitem{CKP13} Contopoulos, I., Kazanas, D. \& Papadopoulos, D.B. 2013, ApJ,{\bf765}113
\bibitem{GJ14} Gralla, S. E. \& Jacobson, T. 2014, arXiv:1401,6159
\bibitem{GL13} Globus, N. \& Levinson, A. 2013 Phys.Rev. D, {\bf 88}, 4046
\bibitem{KHKC12} Kalapotharakos, C., Harding, A. K., Kazanas, D.\& Contopoulos, I. 2012, ApJ, {\bf 754}, 1
\bibitem{KLK75} King, A. R., Lasota, J. P. \& Kundt, W. 1975,
PThPh, {\bf 12}, 3037
\bibitem{K01} Komissarov, S. S. 2001, MNRAS, {\bf 326}, 41
\bibitem{K04a} Komissarov, S. S. 2004, MNRAS, {\bf 350}, 427 (a)
\bibitem{K04b} Komissarov, S. S. 2004, MNRAS, {\bf 350}, 1431 (b)
\bibitem{KMcK07} Komissarov, S. S. \& McKinney, J. C. 2007, MNRAS, {\bf 377}, L49
\bibitem{LGATN} Lasota, J. P., Gourgoulhon, E., Abramowicz, M.,
Tchekhovskoy, A. \& Narayan, R. 2014 Phys.Rev. D, 89, 024041
\bibitem{LST12b} Li, J., Spitkovsky, A. \& Tchekhovskoy, A.2012,ApJ, {\bf 746}, 60
\bibitem{LK01} Lyubarsky, Y. \& Kirk, J. G. 2001, ApJ, {\bf547},437
\bibitem{L10} Lyutikov, M. 2010, arXiv:1010.6254
\bibitem{L12} Lyutikov, M. 2012, arXiv:1209.3785
\bibitem{k3} MacDonald, D. A. \& Thorne, K. S. 1982, MNRAS, {\bf 198}, 345
\bibitem{McG04} McKinney, J. C., Gammie, C. F. 2004, ApJ, {\bf 611}, 977
\bibitem{M82} Michel, F. C. 1982, Rev. Mod. Phys., {\bf 54}, 1
\bibitem{NMcC12} Narayan, R. \& McClintock, J. E. 2012, MNRAS, {\bf 419}, 69
\bibitem{OT03} Ogura, J. \& Kojima, Y. 2003, PThPh, {\bf 109}, 619
\bibitem{P10} Palenzuela, C., Garrett, T., Lehner, L. \&
Liebling, S. L. 2010, Phys. Rev. D, {\bf 82}, 044045
\bibitem{P14} Pena, R. 2014, arXiv:1403.0938
\bibitem{PFTV86} Press, W. H., Flannery, B. P., Teukolsky, S. A. 1986, `Numerical
recipes. The art of scientific computing', Cambridhe Univ. Press
\bibitem{PC90} Punsly, B. \& Coroniti, F. V. 1990, ApJ, {\bf 354}, 583
\bibitem{RGF13} Russell, D. M., Gallo,E. \& Fender, R. P. 2013, MNRAS, {\bf 431}, 405
\bibitem{SNPZ13} Sadowski, A., Narayan, R., Pena, R. \& Zhu, Y. 2013, MNRAS, {\bf
436}, 3856
\bibitem{SW73} Scharlemann, E. T. \& Wagoner, R. V. 1973, ApJ, {\bf 182}, 951
\bibitem{SB13} Sikora, M. \& Begelman, M. C. 2013,
ApJ, {\bf 764}, L24
\bibitem{S07} Sikora, M., Stawarz, L., Lasota, J. P. 2007,
ApJ, {\bf 658}, 815
\bibitem{TNMcK10} Tchekhovskoy, A., Narayan, R. \& McKinney, J. C. 2010, ApJ, {\bf711},50
\bibitem{TPM} Thorne, K. S., Price, R. H. \& MacDonald D. A. {\em Black Holes:
The Membrane Paradigm} Yale University Press 1986
\bibitem{U05} Uzdensky, D. A. 2005, ApJ, {\bf 620}, 889
\bibitem{W74} Wald, R. M. 1974, Phys. Rev. D, {\bf 10}, 1680
\bibitem{Z77} Znajek, R. L. 1977, MNRAS, {\bf 179}, 457


\end{thebibliography}
\end{document}